\documentclass[onecolumn]{revtex4}

\usepackage{graphicx}
\usepackage{dcolumn}
\usepackage{bm}

\input epsf

\begin{document}

\title{\Large Study of Thermodynamic Quantities in Generalized Gravity Theories}

\author{\bf
Surajit Chattopadhyay$^1$\footnote{surajit$_{-}$2008@yahoo.co.in},
Ujjal Debnath$^2$\footnote{ujjaldebnath@yahoo.com ,
ujjal@iucaa.ernet.in} and Samarpita
Bhattacharya$^2$\footnote{samarpita$_{-}$sarbajna@yahoo.co.in}}

\affiliation{$^1$Department of Computer Application (Mathematics
Section), Pailan College of Management and Technology, Bengal
Pailan Park,
Kolkata-700 104, India.\\
$^2$Department of Mathematics, Bengal Engineering and Science
University, Shibpur, Howrah-711 103, India.}

\date{\today}

\begin{abstract}
In this work, we have studied the thermodynamic quantities like
temperature of the universe, heat capacity and squared speed of
sound in generalized gravity theories like Brans-Dicke,
Ho$\check{\text r}$ava-Lifshitz and $f(R)$ gravities. We have
considered the universe filled with dark matter and dark energy.
Also we have considered the equation of state parameters for open,
closed and flat models. We have observed that in all cases the
equation of state behaves like quintessence. The temperature and
heat capacity of the universe are found to decrease with the
expansion of the universe in all cases. In Brans-Dicke and $f(R)$
gravity theories the squared speed of sound is found to exhibit
increasing behavior for open, closed and flat models and in
Ho$\check{\text r}$ava-Lifshitz gravity theory it is found to
exhibit decreasing behavior for open and closed models with the
evolution of the universe. However, for flat universe, the squared
speed of sound remains constant in Ho$\check{\text r}$ava-Lifshitz
gravity.
\end{abstract}


\maketitle

\section{\normalsize\bf{Introduction}}

Recently, it has become well known that the universe has not only
undergone the period of early-time accelerated expansion
(inflation), but also is currently in the so-called late-time
accelerating epoch (dark energy era). The unified description of
inflation and dark energy is achieved by modifying the
gravitational action at the very early Universe as well as at the
very late times [1, 2]. A number of viable modified gravity
theories has been suggested [3, 4, 5, 6, 7]. In reference [8], the
connection between modified gravity and M-string theory was
indicated. The modified gravity gives the qualitative answers to
the number of fundamental questions about dark energy. Indeed, the
origin of dark energy may be explained by some sub-leading
gravitational terms which become relevant with the decrease of the
curvature (at late times). Moreover, there are many proposals to
consider the gravitational terms relevant at high curvature
(perhaps, due to quantum gravity effects) as the source of the
early-time inflation. Hence, there appears the possibility to
unify and to explain both: the inflation and late-time
acceleration as the modified gravity effects [8]. Reviews on
modified gravity are available in the references like [9] and
[10]. Among the recent attempts to construct a consistent theory
of quantum gravity, much attention has been paid to the quite
remarkable Ho$\check{\text r}$ava-Lifshitz quantum gravity [11].
An extensively studied generalization of general relativity
involves modifying the Einstein-Hilbert Lagrangian in the simplest
possible way, replacing $R-2\Lambda$ by a more general function
$f(R)$ [10, 11, 12]. Recently the modified Ho$\check{\text
r}$ava-Lifshitz $f(R)$ gravity has been proposed in ref.[1].
Discussions on Ho$\check{\text r}$ava-Lifshitz gravity have been
made in references [13, 14]. The basic idea of Ho$\check{\text
r}$ava-Lifshitz gravity is to modify the UV behavior of the
general theory so that the theory is perturbatively renormalizable
[14]. However this modification is only possible on condition when
we abandon Lorentz symmetry in the high energy regime [14]. In
reference [15], the very interesting physical implications of
Ho$\check{\text r}$ava-Lifshitz gravity are summarized as: (i) the
novel solution subclasses, (ii) the gravitational wave production,
the perturbation spectrum, (iii) the matter bounce, (iv) the dark
energy phenomenology, (iv) the astrophysical phenomenology, and
(v) the observational constraints on the theory. Recently,
scalar-tensor theories have received renewed interest. The
Brans-Dicke theory [16] is the simplest example of a scalar-tensor
theory of gravity. In Brans-Dicke theory, Newton's constant
becomes a function of space and time, and a new parameter $\omega$
is introduced. General relativity is recovered in the limit
$\omega\rightarrow\infty$ [17]. Interacting dark energy [18, 19]
and holographic dark energy [20, 21, 22, 23, 24] models have been
considered in Brans-Dicke theory. Brans-Dicke scalar field as
chameleon field has been considered in the references [25] and [26].\\

A profound connection between gravity and thermodynamics was first
established by Jacobson [27], who first showed that the Einstein
gravity can be derived from the first law of thermodynamics in the
Rindler spacetime. Thermodynamic aspects of the cosmological
horizons have been reviewed in [28] and [29]. Investigating the
generalized second law (GSL) of thermodynamics in gravity has
gained immense interest in recent years. A plethora of papers have
studied the thermodynamics in Einstein gravity theory [30, 31, 32,
33, 34, 35]. As the modified theory of gravity was argued to be a
possible candidate to explain the accelerated expansion of our
universe by various authors [36, 37, 38], thus it is interesting
to examine the GSL in the extended gravity theories [39, 40, 41,
42]. Thermodynamics has been studied in the brane world scenario
[43, 44, 45, 46],  Ho$\check{\text r}$ava-Lifshitz gravity [47,
48, 49], Brans-Dicke gravity [50, 51, 52] and in $f(R)$ gravity
[53, 54, 55]. Extending the study of [15], two of the authors of
the present paper, examined the validity of the GSL in various
cosmological horizons of a universe governed by the
Ho$\check{\text r}$ava-Lifshitz gravity and the GSL was proved to
be valid in different horizons [56].\\

In the present work, we have studied the thermodynamic quantities
of the universe in generalized gravity theories like
 Brans-Dicke, Ho$\check{\text r}$ava-Lifshitz and $f(R)$ gravities. Instead of investigating
the validity of the laws of thermodynamics, we have tried to
investigate how the thermodynamic quantities like heat capacity
$(C_{v})$, temperature $T$ and squared speed of sound $v_{s}^{2}$
behave during the evolution of the universe governed by the said
gravity theories. In addition to this, the equation of state
parameters have also been studied for all of the said gravity
theories. Organization of the rest of the paper is as follows: In
section II, we have discussed the thermodynamic quantities. In
sections IIIA, IIIB and IIIC we have discussed the thermodynamic
quantities under Brans-Dicke, Ho$\check{\text r}$ava-Lifshitz and
$f(R)$ gravity theories respectively. Finally, in section IV, we
have discussed the results.\\

\section{\normalsize\bf{General Description of Thermodynamic Quantities}}

 The Einstein field equations for homogeneous, isotropic FRW
universe are given by [57] (choosing $c=1$)

\begin{equation}
{H}^{2}+\frac{k}{a^{2}}= \frac{8\pi G}{3}\rho
\end{equation}

and

\begin{equation}
\dot{H}-\frac{k}{a^{2}}=-4\pi G(\rho+p)
\end{equation}

where $H(=\frac{\dot{a}}{a}$) is the Hubble parameter and
$k=0,-1,+1$ denote the curvature index for flat, open and closed
universe respectively. Here, $\rho$ and $p$ denote the energy
density and pressure of the universe. The energy momentum tensor
$T_{\nu}^{\mu}$ is conserved by virtue of the Bianchi identities,
leading to the continuity equation [57]

\begin{equation}
\dot{\rho} + 3 H (\rho + p) =0
\end{equation}

where $p$ is the isotropic pressure and $\rho$ is the energy
density of the fluid defined by

\begin{equation}
\rho=\frac{U}{V}
\end{equation}

Here, $U$ is the internal energy and $V$ is the volume of the
universe.\\

We consider the FRW universe treated as a thermodynamical system.
Then from Gibb's equation of thermodynamics, we have [33]

\begin{equation}
TdS=d(\rho V)+ pdV = d((\rho +p)V)-Vdp
\end{equation}

where $S$ is the entropy, $T$ is the temperature and $V$ is the
volume of the universe. The integrability condition of
thermodynamic system is given by [58]

\begin{equation}
\frac{\partial^{2}S}{\partial T \partial V} = \frac{\partial^{2}S
} {\partial V \partial T }
\end{equation}

which leads to the relation between pressure, energy density and
temperature as

 \begin{equation}
dp= \frac{\rho + p}{T} dT
\end{equation}

From (5) and (7), we get

\begin{equation}
dS=d\left(\frac{( \rho + p )V}{T}\right)
\end{equation}

and integrating, we can obtain the  expression of the entropy as
(except for an additive constant)

\begin{equation}
S=\frac{(\rho+p) V}{T}
\end{equation}

However, for adiabatic process entropy is constant and
consequently, the equation (5) becomes

\begin{equation}
d[(\rho + p)]=Vdp
\end{equation}

Relation $(9)$ can also be obtained using $(7)$ into $(10)$. Hence
for adiabatic process equation $(9)$ may be considered as the
temperature defining equation as

\begin{equation}
T=\frac{(\rho + p)V}{S}
\end{equation}

The square speed of sound and heat capacity are defined by

\begin{equation}
v_{s}^{2}=\frac{\partial p}{\partial \rho}
\end{equation}
and
\begin{equation}
C_{_{V}}=V\frac{\partial\rho}{\partial T} ~~\text{(when entropy
$S$ is constant = $S_{0}$, say)}
\end{equation}

These thermodynamic quantities would be investigated for their
evolution with the expansion of the universe in the subsequent
sections.\\

\section{\normalsize\bf{thermodynamic quantities in Generalized Gravity Theories}}

\subsection{\normalsize\bf{Brans-Dicke Theory}}

The Jordan-Fierz-Brans-Dicke theory (heretofore, we will call it
Brans-Dicke (BD) theory for simplicity) is the simplest example of
a scalar-tensor theory of gravity. A brief introduction of the BD
theory has been presented in the previous section. The Lagrangian
density for the Brans-Dicke theory is

\begin{equation}
\mathcal{L}=\sqrt{-g}\left[-\phi
R+\frac{\omega}{\phi}g^{\mu\nu}\partial_{\mu}\phi+\mathcal{L}_{m}\right]
\end{equation}
where $\phi$ is the Brans-Dicke field, and $\mathcal{L}_{m}$ is
the Lagrangian density for the matter fields. The self-interacting
BD theory is described by the Jordan-Brans-Dicke (JBD) action
(choosing $c=1$) as:

\begin{equation}
S=\int\frac{d^{4} x \sqrt{-g}}{16\pi}\left[\phi R-
\frac{\omega(\phi)}{\phi} {\phi}^{,\alpha}
{\phi,}_{\alpha}-V(\phi)+ 16\pi{\cal L}_{m}\right]
\end{equation}

where $V(\phi)$ is the self-interacting potential for the BD
scalar field $\phi$ and $\omega(\phi)$ is modified version of the
BD coupling parameter which is a function of $\phi$. In this
theory $\frac{1}{\phi}$ plays the role of the gravitational
constant $G$. This action also matches with the low energy string
theory action for $\omega=-1$. The matter content of the Universe
is composed of matter fluid, so the energy-momentum tensor is
given by

\begin{equation}
T_{\mu \nu}^{m}=(\rho+p)u_{\mu} u_{\nu}+p~g_{\mu \nu}
\end{equation}

where $u^{\mu}$ is the four velocity vector of the matter fluid
satisfying $u_{\mu}u^{\nu}=-1$ and $\rho,~p$ are respectively
energy density and isotropic pressure.\\

From the Lagrangian density we obtain the field equations

\begin{equation}
G_{\mu \nu}=\frac{8\pi}{\phi}T_{\mu
\nu}^{m}+\frac{\omega(\phi)}{{\phi}^{2}}\left[\phi  _{ , \mu} \phi
_{, \nu} - \frac{1}{2}g_{\mu \nu} \phi _{, \alpha} \phi ^{ ,
\alpha} \right]+\frac{1}{\phi}\left[\phi  _{, \mu ; \nu} -g_{\mu
\nu}~ ^{\fbox{}}~ \phi \right]-\frac{V(\phi)}{2 \phi} g_{\mu \nu}
\end{equation}

and

\begin{equation}
^{\fbox{}}~\phi=\frac{8\pi
T}{3+2\omega(\phi)}-\frac{1}{3+2\omega(\phi)}\left[2V(\phi)-\phi
 \frac{dV(\phi)}{d\phi}\right]
-\frac{\frac{d\omega(\phi)}{d\phi}}{3+2\omega(\phi)}{\phi,}_{\mu}
 {\phi}^{,\mu}
\end{equation}

where $T=T_{\mu \nu}^{m}g^{\mu \nu}$. Equation (17) can also be
written as

\begin{equation}
G_{\mu\nu}=8\pi
\tilde{T}_{\mu\nu}=\frac{8\pi}{\phi}\left(T_{\mu\nu}^{m}+\frac{1}{8\pi}T_{\mu\nu}^{\phi}\right)
\end{equation}

where $\tilde{T}_{\mu\nu}$ can be treated as effective energy
momentum tensor. The line element for Friedman-Robertson-Walker
space-time is given by

\begin{equation}
ds^{2}=-dt^{2}+a^{2}(t)\left[\frac{dr^{2}}{1-kr^{2}}+r^{2}(d\theta^{2}+sin^{2}\theta
d\phi^{2})\right]
\end{equation}

where, $a(t)$ is the scale factor and $k~(=0, -1,+1)$ is the
curvature index describe the flat, open and closed model of the
universe.\\

We are considering the universe filled with dark energy (with
energy density $\rho_{D}$) and dark matter (with energy density
$\rho_{m}$). As we are not considering interacting situation, the
conservation equations are separately satisfied for dark matter
and dark energy. Thus

\begin{equation}
\dot{\rho}_{D}+3H(\rho_{D}+p_{D})=0
\end{equation}
and
\begin{equation}
\dot{\rho}_{m}+3H(1+w_{m})\rho_{m}=0
\end{equation}

Solving the conservation equation for dark matter, we get the
density and pressure of dark matter as $\rho_{m}=\rho_{m_{0}}(1 +
z) ^{3(1 + w_{m})}$ and $p_{m}=\rho_{m_{0}}(1 + w_{m})(1 + z)^{3(1
+ w_{m})}$. Defining $\rho_{1}=\frac{\rho_{m}}{\phi}$ and
$p_{1}=\frac{p_{m}}{\phi}$ the Einstein's field equations can be
written as

\begin{equation}
H^{2}+\frac{k}{a^{2}}=\frac{8 \pi}{3}(\rho_{1} + \rho_{D})
 \end{equation}

 \begin{equation}
\dot{H}- \frac{k}{ a^{2}}=-4 \pi(\rho_{1}+p_{1}+\rho_{D} + p_{D})
\end{equation}

\begin{figure}
\includegraphics[height=2.8in]{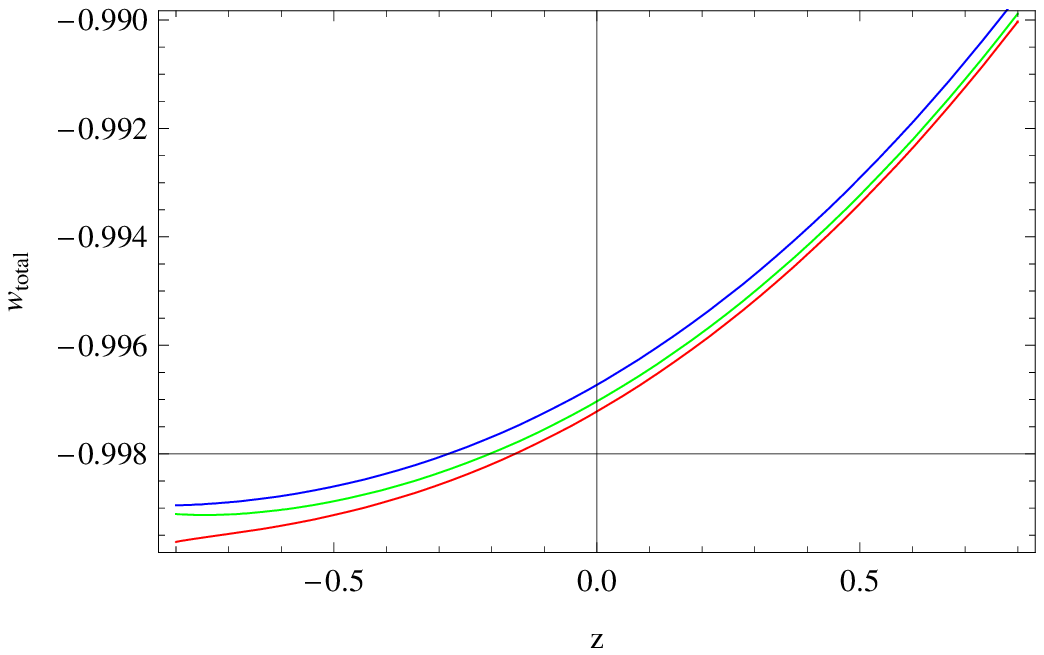}\\
\vspace{1mm}~~~~~~~Fig.1\\ \vspace{6mm} Fig. 1 shows the EOS
parameter $w_{total}=\frac{p_{1}+p_{D}}{\rho_{1}+\rho_{D}}$ for
$k=-1$ (the red line), $k=1$ (the green line) and $k=0$ (the blue
line) for Brans-Dicke model where dark energy and dark matter
satisfy the conservation equation separately. We have taken $\alpha=3$, $B=0.1$, $\rho_{m0}=0.23$, $w_{m}=0.003$.
\\

\vspace{6mm}

\end{figure}

\begin{figure}
\includegraphics[height=2.1in]{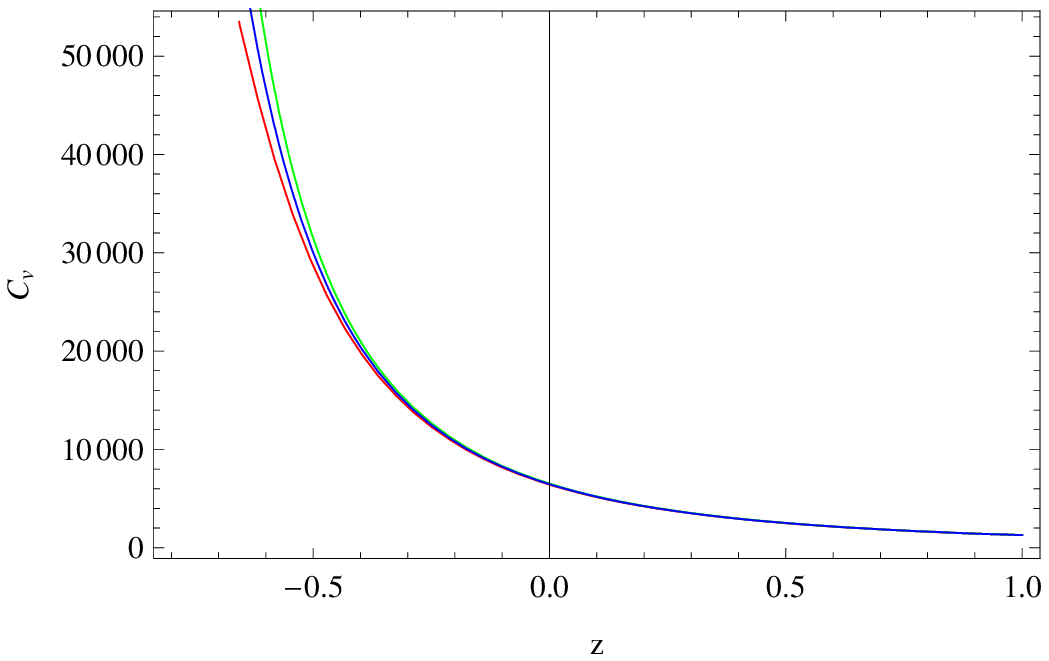}~~~
\includegraphics[height=2.1in]{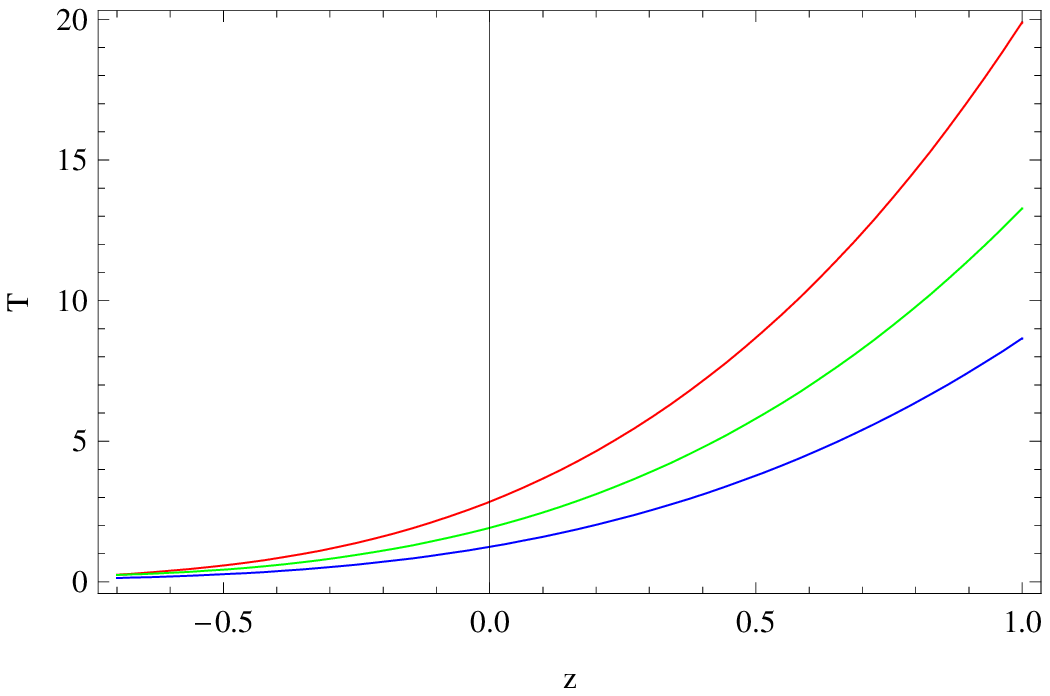}\\
\vspace{1mm} ~~~~~~~~~~~~Fig.2~~~~~~~~~~~~~~~~~~~~~~~~~~~~~~~~~~~~~~~~~~~~~~~~~~Fig.3\\
\vspace{6mm}
\includegraphics[height=2.1in]{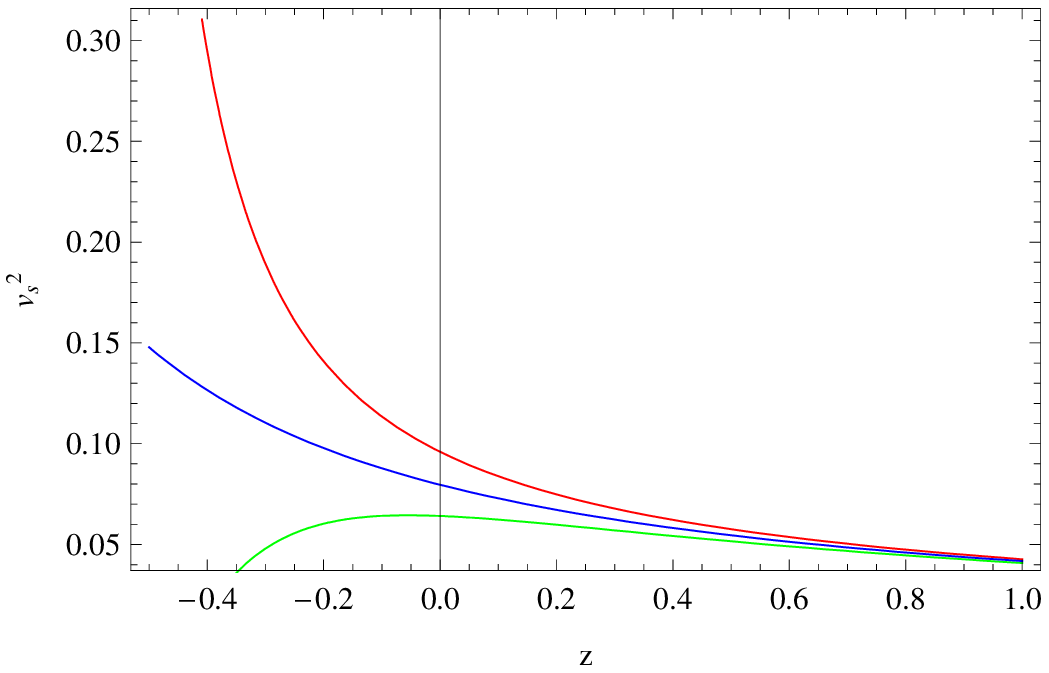}\\
\vspace{6mm} ~~~~~~~~~~~~Fig.4\\
 Fig. 2 shows the heat capacity
$C_{v}$ for $k=-1$ (the thick line), $k=1$ (the dotted line) and
$k=0$ (the broken line) for Brans-Dicke model where dark energy
and dark matter satisfy
the conservation equation separately. We have taken $\alpha=3$, $B=0.1$, $\rho_{m0}=0.23$, $w_{m}=0.003$. \\
Fig. 3 shows the temperature $T$ with evolution of the universe
for $k=-1$ (the red line), $k=1$ (the green line) and
$k=0$ (the blue line) for Brans-Dicke model. We have taken $\alpha=3$, $B=0.1$, $\rho_{m0}=0.23$, $w_{m}=0.003$.\\
Fig. 4 shows the squared speed of sound $v_{s}^{2}$ with evolution
of the universe for $k=-1$ (the red line), $k=1$ (the green line)
and
$k=0$ (the blue line) for Brans-Dicke model. We have taken $\alpha=3$, $B=0.1$, $\rho_{m0}=0.23$, $w_{m}=0.003$.\\

\vspace{6mm}

\end{figure}

where

\begin{equation}
\rho_{D}= \frac{\omega}{16 \pi } \frac{\dot{\phi}^{2}}{\phi ^{2}}
- \frac{3}{8 \pi}H \frac{\dot{\phi}}{\phi} + \frac{V(\phi)}{16 \pi
\phi}
\end{equation}
and
\begin{equation}
p_{D}= \frac{\omega}{16 \pi} \frac{\dot{\phi}^{2}}{\phi ^{2}} +
\frac{H}{4 \pi} \frac{\dot{\phi}}{\phi} + \frac{1}{8 \pi}
\frac{\ddot{\phi}}{\phi} - \frac{V(\phi)}{16 \pi \phi}
\end{equation}

To find the thermal quantities we use the choices of $\phi$ and
$V$ as
\begin{equation}
\phi=\phi_{0}a^{\alpha},~~~~~~~~~ V=V_{0} \phi^{\frac{-3 (1 +
w_{m})}{\alpha}}
\end{equation}

Using EOS for dark energy $p_{D}=w\rho_{D}$ and the solution of
(21) $\rho_{D}=\rho_{D0}a^{-3(1+w)}$ in the field equations we get

\begin{equation}
H^{2}=k A a^{-2}+B
a^{-\frac{2\alpha((1+\omega)\alpha-1)}{2+\alpha}}+Ca^{-\alpha-3(1+w)}
\end{equation}

\begin{equation}
\dot{H}=k A_{1} a^{-2}+B_{1}
a^{-\frac{2\alpha((1+\omega)\alpha-1}{2+\alpha}}+C_{1}a^{-\alpha-3(1+w)}
\end{equation}

where,

\begin{equation}
\begin{array}{c}
  A=\frac{2}{\alpha((1+\omega)\alpha-1)-(2+\alpha)};~~~C=-\frac{16\pi(1+w)\rho_{D0}}{\phi_{0}[(2+\alpha)(-\alpha-3(1+w))+2\alpha((1+\omega)\alpha-1)]} \\
  A_{1}=\frac{2+\alpha(1-\alpha(1+\omega))A}{2+\alpha};~~~ B_{1}=\frac{\alpha(1-\alpha(1+\omega))B}{2+\alpha};~~~C_{1}=\frac{\alpha(1-\alpha(1+\omega))\phi_{0}C-8\pi(1+w)\rho_{0}
}{(2+\alpha)\phi_{0}}\\
   A_{2}=6\phi_{0}+AB_{2}/B;~~~ B_{2}=6\rho_{D0}(1+\alpha-\omega\alpha^{2}/6)B;~~~ C_{2}=B_{2}C/B-16\pi\rho_{D0}\\
\end{array}
\end{equation}

Equation of state parameter
$w_{total}=\frac{p_{1}+p_{D}}{\rho_{1}+\rho_{D}}$ is computed for
flat, open as well as closed universes and are plotted against
redshift $z$ in figure 1 and it is found that in all of the above
cases the EOS parameters are staying above $-1$, which indicates
quintessence-like behavior. The behaviour of the EOS parameters
further indicate that the energy density is increasing with
evolution of the universe irrespective of it curvature.\\

We replace $p$ and $\rho$ by $(p_{1}+p_{D})$ and
$(\rho_{1}+\rho_{D})$ in equation (11) we get the temperature $T$,
which is used in equation (13) to get the heat capacity $C_{v}$.
Heat capacity $C_{v}$ is computed for open, closed and flat
universes and are plotted in figure 2. This figure shows that for
all of the three universes, the heat capacity is decreasing with
increase in the redshift. This means that the heat capacity is
increasing with evolution of the universe irrespective of its
curvature. Also, we present the temperature $T$ against redshift
$z$ for all of the curvatures. We find that the temperature is
decreasing with evolution of the universe.\\

\subsection{\normalsize\bf{Ho$\check{\text r}$ava-Lifshitz
Gravity}}

Thermodynamics in cosmology has been extensively studied either in
Einstein's theory of gravity or in modified theories of gravity.
In this section, we shall generalize such studies to the
Ho$\check{\text r}$ava-Lifshitz (HL) Cosmology. An exhaustive
review of HL cosmology is available in [59]. We briefly review the
scenario where the cosmological evolution is governed by HL
gravity. The dynamical variables are the lapse and shift
functions, $N$ and $N_{i}$ respectively, and the spatial metric
$g_{ij}$. In terms of these fields the full metric is written as
[60]

\begin{equation}
ds^{2}=-N^{2}dt^{2}+g_{ij}(dx^{i}+N^{i}dt)(dx^{j}+N^{j}dt)
\end{equation}

where indices are raised and lowered using $g_{ij}$. The scaling
transformation of the coordinates reads: $t\rightarrow l^{3}t$ and
$x^{i}\rightarrow lx^{i}$.

The action of the HL gravity is given by [60]

\begin{equation}
\begin{array}{c}
  I=dt
  \int dtd^{3}x(\mathcal{L}_{0}+\mathcal{L}_{1}+\mathcal{L}_{m})\\
  \mathcal{L}_{0}=\sqrt{g}N \left[\frac{2}{\kappa^{2}}(K_{ij}K^{ij}-\lambda K^{2})+\frac{\kappa^{2}\mu^{2}(\Lambda R-3\Lambda^{2})}{8(1-3 \lambda)}\right]\\
  \mathcal{L}_{1}=\sqrt{g}N\left[\frac{\kappa^{2}\mu^{2}(1-4\lambda)}{32(1-3\lambda)}R^{2}-\frac{\kappa^{2}}{2\omega^{4}}(C_{ij}-\frac{\mu\omega^{2}}{2}R_{ij})(C^{ij}-\frac{\mu\omega^{2}}{2}R^{ij})\right]
  \\
\end{array}
\end{equation}

where, $\kappa^{2}$, $\lambda$, $\mu$, $\omega$ and $\Lambda$ are
constant parameters, and $C_{ij}$ is Cotton tensor (conserved and
traceless, vanishing for conformally flat metrics). The first two
terms in $\mathcal{L}_{0}$ are the kinetic terms, others in
$(\mathcal{L}_{0} +\mathcal{L}_{1})$ give the potential of the
theory in the so-called ``detailed-balance" form, and
$\mathcal{L}_{m}$ stands for the Lagrangian of other matter field.
Comparing the action to that of the general relativity, one can
see that the speed of light and the $cosmological$ Newton's
constant are

\begin{figure}
\includegraphics[height=2.8in]{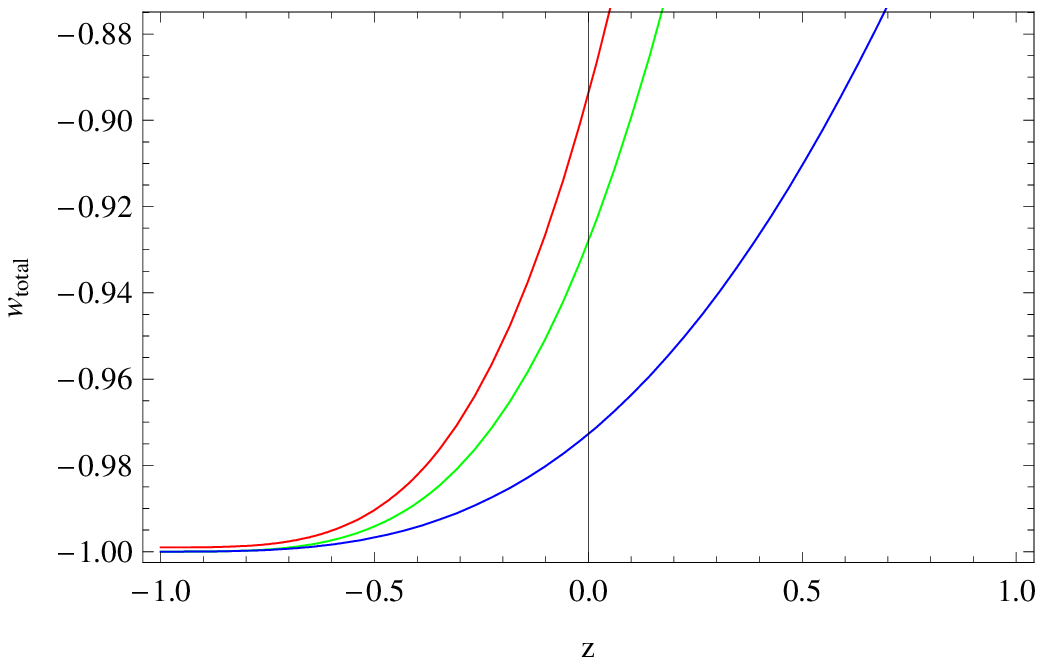}~~~\\
\vspace{1mm} ~~~~~~~~~~~~Fig.5\\
\vspace{6mm} Fig. 5 shows the behavior of the the equation of
state parameter $w_{total}$ with evolution of the universe for
$k=-1$ (the red line), $k=1$ (the green line) and $k=0$ (the blue
line) for Ho$\check{\text r}$ava-Lifshitz gravity. We have taken
$\lambda=1.2$, $\mu=1.03$, $w_{m}=0.03$ and $\rho_{m0}=0.23$.

\vspace{6mm}

\end{figure}

\begin{figure}
\includegraphics[height=2.1in]{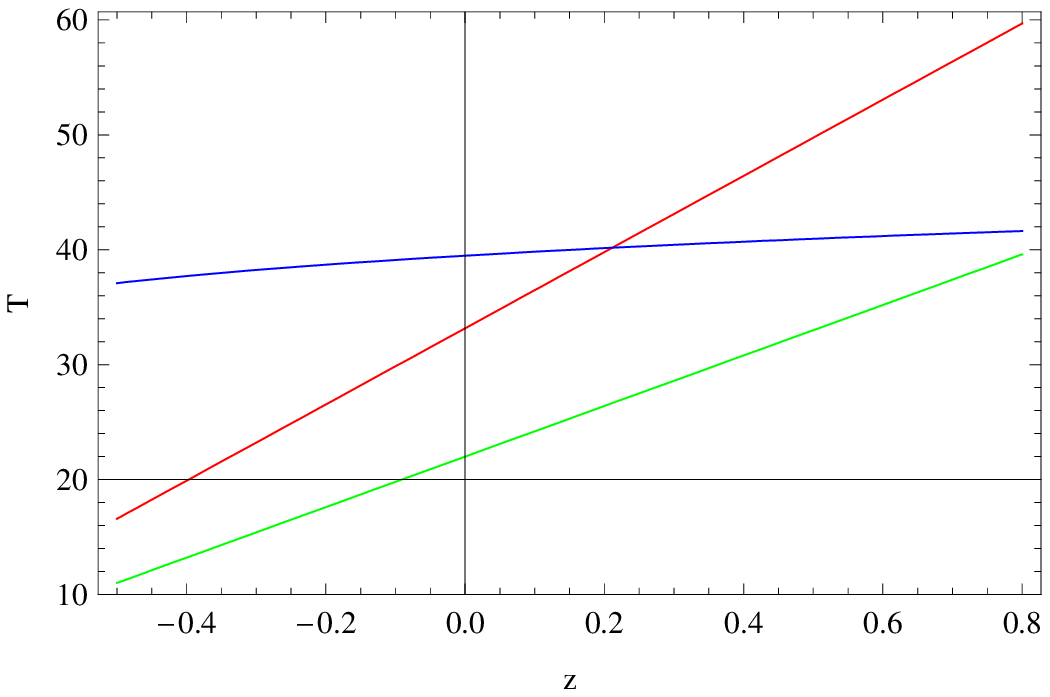}~~~
\includegraphics[height=2.1in]{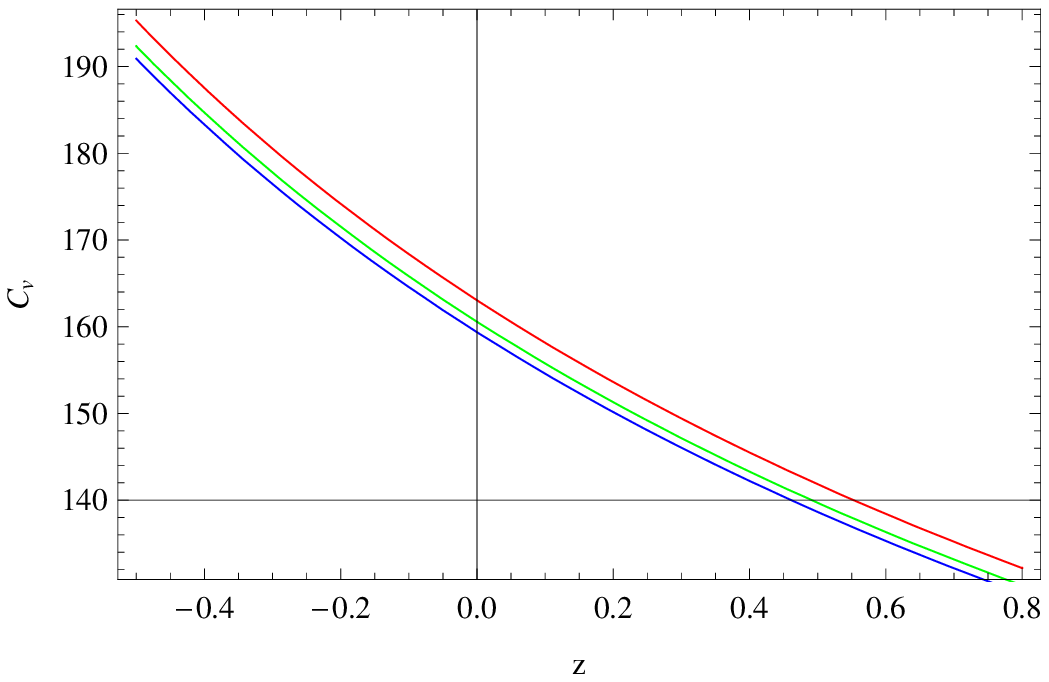}]\\
\vspace{1mm} ~~~~~~~~~~~~Fig.6~~~~~~~~~~~~~~~~~~~~~~~~~~~~~~~~~~~~~~~~~~~~~~~~~~~~~~~~~~~~~~Fig.7\\
\vspace{6mm} Fig. 6 shows the behavior of temperature $T$ with
evolution of the universe for $k=-1$ (the red line), $k=1$ (the
green line) and $k=0$ (the blue line) for Ho$\check{\text
r}$ava-Lifshitz gravity. We have taken
$\lambda=1.2$, $\mu=1.03$, $w_{m}=0.03$ and $\rho_{m0}=0.23$.\\
Fig. 7 shows the behavior of heat capacity $C_{v}$ with evolution
of the universe for $k=-1$ (the red line), $k=1$ (the green line)
and $k=0$ (the blue line) for Ho$\check{\text r}$ava-Lifshitz
gravity. We have taken $\lambda=1.2$, $\mu=1.03$, $w_{m}=0.03$ and
$\rho_{m0}=0.23$.

\vspace{6mm}

\end{figure}

\begin{figure}
\includegraphics[height=2.1in]{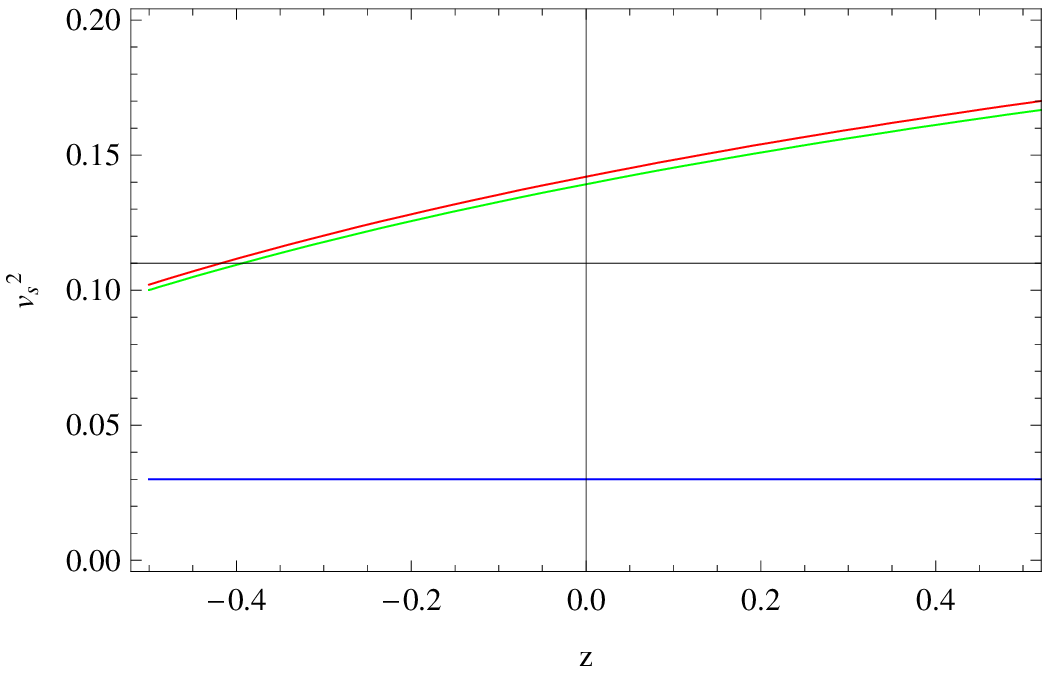}~~~\\
\vspace{1mm} ~~~~~~~~~~~~Fig.8\\
\vspace{6mm} Fig. 8 shows the squared speed of sound $v_{s}^{2}$
with evolution of the universe for $k=-1$ (the red line), $k=1$
(the green line) and $k=0$ (the blue line) for Ho$\check{\text
r}$ava-Lifshitz gravity. We have taken $\lambda=1.2$, $\mu=1.03$,
$w_{m}=0.03$ and $\rho_{m0}=0.23$.

\vspace{6mm}

\end{figure}

\begin{equation}
c=\frac{\kappa^{2}\mu}{4}\sqrt{\frac{\Lambda}{1-3\lambda}},~~~~~G_{c}=\frac{\kappa^{2}c}{16\pi(3\lambda-1)}
\end{equation}

It may be noted that when $\lambda=1$, $\mathcal{L}_{0}$ reduces
to the usual Lagrangian of Einstein's general relativity. Thus,
when $\lambda=1$, the general relativity is approximately
recovered at large distances.\\

As we are considering dark energy with dark matter the
conservation equations are given by (21) and (22). The field
equations are

\begin{equation}
H^{2} + \frac{k}{a^{2}}=\frac{8\pi G_{c}}{3}( \rho_{m} + \rho_{D})
\end{equation}
and
\begin{equation}
\dot{H}+\frac{3}{2}H^{2}+\frac{k}{2a^{2}}= -4 \pi
G_{c}\left(p_{m}+ p_{D}\right)
\end{equation}

where

\begin{equation}
\rho_{D}\equiv \frac{3 \kappa^{2} \mu^{2} k^{2}}{8(3\lambda
-1)a^{4}} + \frac{3 \kappa^{2} \mu^{2} \Lambda^{2}}{8(3\lambda
-1)}\equiv\frac{1}{16 \pi G_{c}}\left( \frac{3 k^{2}}{ \Lambda
a^{4}} +  3\Lambda \right)
\end{equation}
and
\begin{equation}
p_{D}\equiv \frac{ \kappa^{2} \mu^{2} k^{2}}{8(3\lambda -1)a^{4}}
- \frac{3 \kappa^{2} \mu^{2} \Lambda^{2}}{8(3\lambda
-1)}\equiv\frac{1}{16 \pi G_{c}}\left( \frac{ k^{2}}{ \Lambda
a^{4}} - 3 \Lambda \right)
\end{equation}

Using the solution for the conservation equation for dark matter
given in (22) we get the total energy density as a function of
redshift $(z=\frac{1}{a}-1)$ as

\begin{equation}
\rho(z)=\rho_{m} +\rho_{D}= \rho_{m_{0}}(1 + z) ^{3(1 + w_{m})} +
\frac{1}{16 \pi G_{c}} \left( \frac{ 3k^{2}(1 + z)^{4}}{ \Lambda }
+ 3\Lambda \right)
\end{equation}

Similarly, the total pressure as a function of redshift $z$ is

\begin{equation}
p(z)=p_{m} + p_{D}=\rho_{m_{0}}w_{m}(1 + z)^{3(1+w_{m})} +
\frac{1}{16 \pi G_{c}} \left( \frac{ k^{2}(1 +z)^{4}}{ \Lambda } -
3 \Lambda \right)
\end{equation}

The squared speed of sound is given as a function of $z$ by

\begin{equation}
v_{s}^{2}(z)=\frac{k^{2}(1+z)\kappa^{2}\mu^{2}+6(1+z)^{3w_{m}}(-1+3\lambda)w_{m}\rho_{m0}(1+w_{m})}{3\{k^{2}(1+z)\kappa^{2}\mu^{2}+2(1+z)^{3w_{m}}(-1+3\lambda)(1+w_{m})\rho_{m0}\}}
\end{equation}

and the heat capacity becomes

\begin{equation}
C_{v}(z)=\frac{3S_{0}\{k^{2}(1+z)\kappa^{2}\mu^{2}+2(1+z)^{3w_{m}}(-1+3\lambda)(1+w_{m})\rho_{m0}\}}{k^{2}(1+z)\kappa^{2}\mu^{2}+6(1+z)^{3w_{m}}(-1+3\lambda)w_{m}\rho_{m0}(1+w_{m})}
\end{equation}

The thermodynamic quantities expressed above are now plotted
against redshift to see  their behavior with the evolution of the
universe. In figure 5, where we have plotted the equation of state
parameter for the Ho$\check{\text r}$ava-Lifshitz gravity, we see
that the behavior is like Brans-Dicke theory. It is staying above
$-1$. However, at lower redshifts the equation of state parameter
is tending to $-1$. However, it never crosses $-1$. Like
Brans-Dicke, this behavior remains the same for flat, open and
closed universes. In figure 6 we find that in the case of flat,
open and closed universes, the temperature $T$ is decreasing with
the evolution of the universe. From figure 6 we see that the heat
capacity $C_{v}$ is increasing as we are approaching towards the
lower redshifts. From figure 7 we understand that for open and
closed universes, the squared speed of sound $v_{s}^{2}$ decreases
with the evolution of the universe. However, for flat universe,
the $v_{s}^{2}$ remains constant throughout the evolution of the
universe.
\\

\subsection{\normalsize\bf{$f(R)$ Gravity}}

Motivated by astrophysical data which indicate that the expansion
of the universe is accelerating, the modified theory of gravity
(or $f(R)$ gravity) which can explain the present acceleration
without introducing dark energy, has received intense attention.
Extensive review of $f(R)$ gravity is available in [61]. The
action of $f(R)$ gravity is given by [62]

\begin{equation}
S=\int
d^{4}x\sqrt{-g}\left[\frac{f(R)}{2\kappa^{2}}+\mathcal{L}_{matter}\right]
\end{equation}
where $g$ is the determinant of the metric tensor $g_{\mu\nu}$,
$\mathcal{L}_{matter}$ is the matter Lagrangian and
$\kappa^{2}=8\pi G$. The$f(R)$ is a non-linear function of the
Ricci curvature $R$ that incorporates corrections to the
Einstein-Hilbert action which is instead described by a linear
function $f(R)$. The gravitational field equations in this theory
are

\begin{equation}
H^{2}+\frac{k}{a^{2}}=\frac{\kappa^{2}}{3f'(R)}(\rho+\rho_{c})
\end{equation}

\begin{equation}
\dot{H}-\frac{k}{a^{2}}=-\frac{\kappa^{2}}{2f'(R)}(\rho+p+\rho_{c}+p_{c})
\end{equation}

where $\rho_{c}$ and $p_{c}$ can be regarded as the energy density
and pressure generated due to the difference of $f(R)$ gravity
from general relativity given by [61] (choosing $G=1$)

\begin{equation}
\rho_{c}=\frac{1}{8 \pi f'}\left[- \frac{f - R f'}{2} -3 H
f''\dot{R}\right]
\end{equation}

\begin{equation}
p_{c}=\frac{1}{8 \pi f'}\left[\frac{f - R f'}{2} + f'' \ddot{R} +
f''' \ddot{R}^{2} + 6 f'' \dot{R}\right]
\end{equation}

where, the scalar tensor $R=-6\left(\dot{H} + 2 H^{2} +
\frac{k}{a^{2}}\right)$.
\\

As we are considering both dark matter and dark energy, the
Friedman equations take the form

\begin{equation}
H^{2}+\frac{k}{a^{2}}=\frac{8 \pi}{3}\rho_{total}
\end{equation}

\begin{equation}
\dot{H} - \frac{k}{ a^{2}}=-4 \pi(\rho_{total} + p_{total})
\end{equation}

where,
\begin{equation}
\rho_{total}=\rho_{1} + \rho_{c},~~~~~~p_{total}= p_{1} + p_{c}
\end{equation}

with $\rho_{1}=\frac{\rho_{m}}{f'}$ and $p_{1}=\frac{p_{m}}{f'}$.
As there is no interaction, like the previous two cases the dark
energy and dark matter satisfy the conservation equation
separately. Therefore, we have the density and pressure of dark
matter as $\rho_{m}=\rho_{m_{0}}(1 + z) ^{3(1 + w_{m})}$ and
$p_{m}=\rho_{m_{0}}(1 + w_{m})(1 + z)^{3(1 + w_{m})}$. In the
present section, while considering the $f(R)$ gravity, we have
illustrated with a solution,

\begin{equation}
f(R)=\beta R + \alpha R^{m},~~~~R=
\frac{A}{a^{n}},~~~m>1,~~\alpha>0,~~\beta>0
\end{equation}
Also we have chosen $n(m - 1)=1$.

\begin{figure}
\includegraphics[height=2.1in]{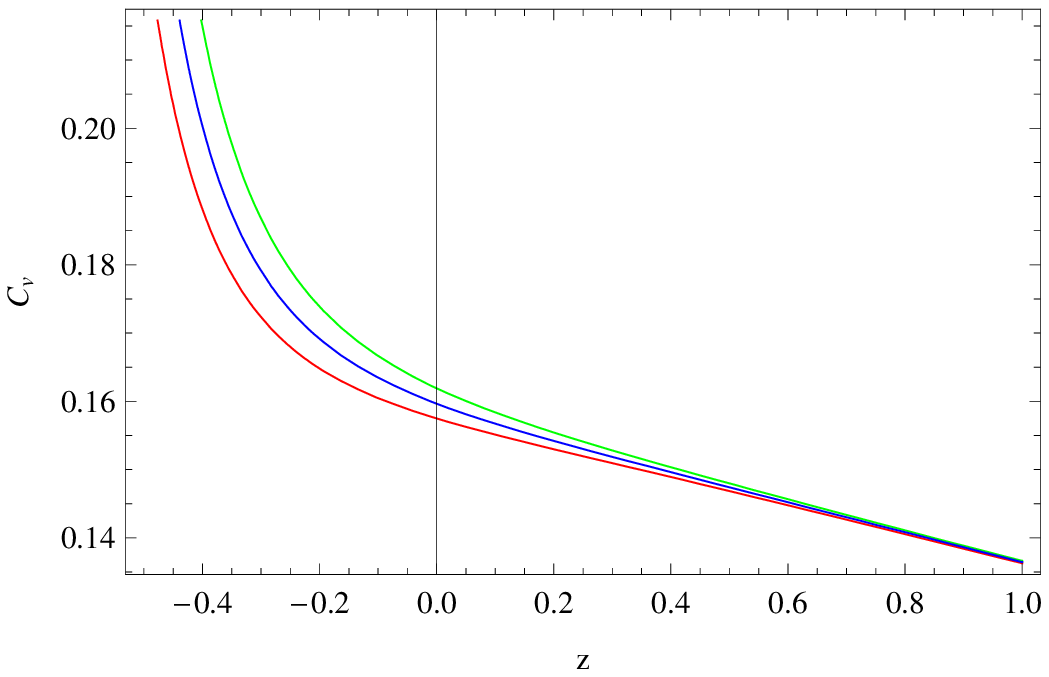}~~~
\includegraphics[height=2.1in]{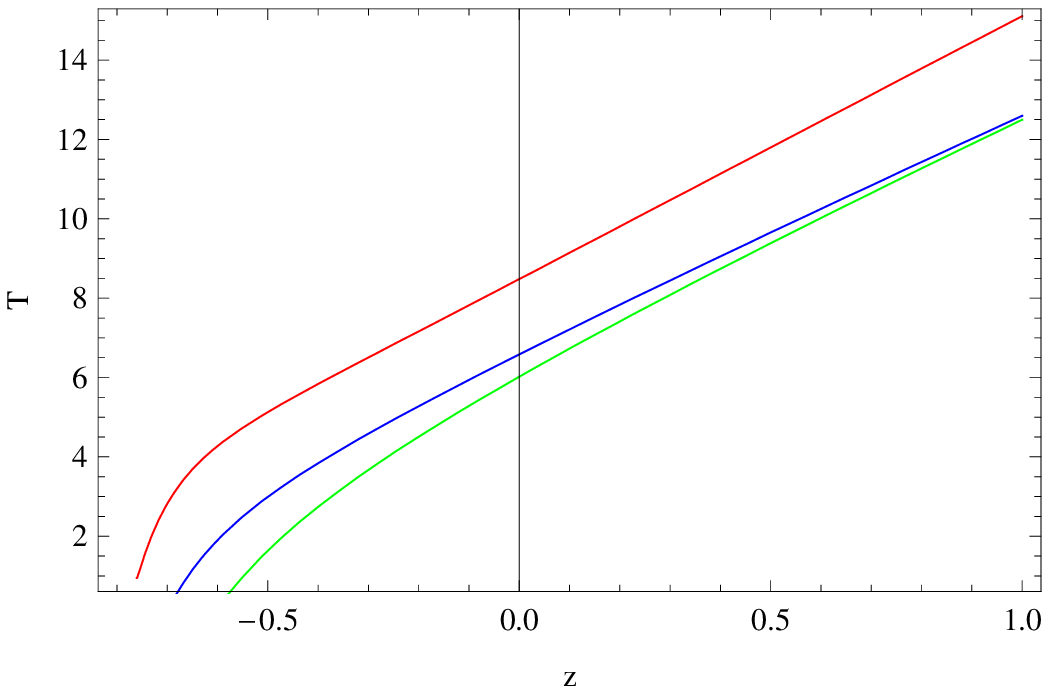}\\
\vspace{1mm} ~~~~~~~~~~~~~~~~~~~Fig.9~~~~~~~~~~~~~~~~~~~~~~~~~~~~~~~~~~~~~~~~~~~~~~Fig.10\\
\vspace{6mm}
\includegraphics[height=2.1in]{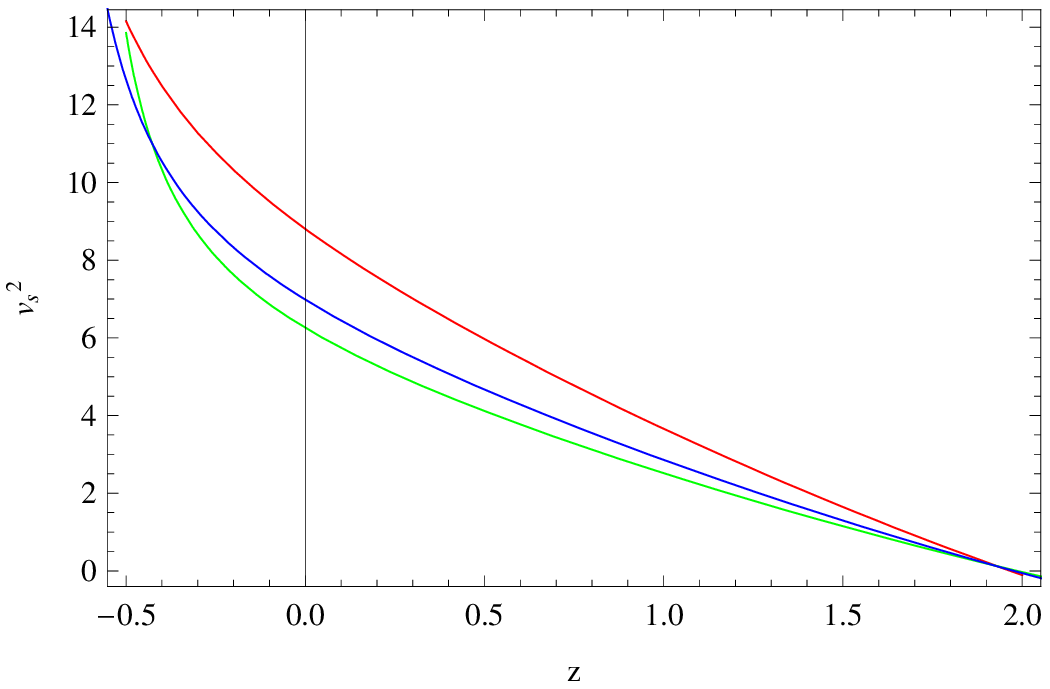}\\
\vspace{1mm} ~~~~~~~~~~~~~~~~~~~Fig.11\\ \vspace{6mm} Fig. 9 shows
the plot of heat capacity $C_{v}$ against redshift $z$ in $f(R)$
gravity. We see that $C_{v}$ is increasing with the evolution of
the universe. Here
$\alpha=2.10132,~~\beta=2.10101,~~w_{m}=0.003,~~m=2.3,~~\rho_{m0}=0.23$
and the red, green and blue lines correspond to $k=-1,~1,~0$ respectively.\\
Fig. 10 shows the plot of temperature $T$ against redshift $z$. We
find that the temperature is decreasing with the evolution of the
universe in $f(R)$ gravity
$\alpha=12.1,~~\beta=10.1,~~w_{m}=0.003,~~m=12.3,~~\rho_{m0}=0.23$
and the red, green and blue lines correspond to $k=-1,~1,~0$ respectively.\\
Fig. 11 shows the plot of the squared speed of sound $v_{s}^{2}$
against redshift $z$ in $f(R)$ gravity. We find that $v_{s}^{2}$
is increasing with the evolution of the universe. Here,
$\alpha=12.1,~\beta=10.1,~w_{m}=0.003,~m=1.9,~\rho_{m0}=0.23$ and
the red, green and blue lines correspond to $k=-1,~1,~0$
respectively.
\\

\vspace{6mm}

\end{figure}

Using the above form of $f(R)$ in (44) and (45) we have computed
the temperature $T$, squared speed of sound $v_{s}^{2}$ and heat
capacity $C_{v}$ as functions of the redshift $z$ as follows:

\begin{equation}
\rho_{1}=\frac{(1+z)^{3(1+w_{m})}\rho_{m0}}{m(A(1+z)^{n})^{-1+m}\alpha+\beta}
\end{equation}

\begin{equation}
p_{1}=\frac{w_{m}(1+z)^{3(1+w_{m})}\rho_{m0}}{m(A(1+z)^{n})^{-1+m}\alpha+\beta}
\end{equation}

\begin{equation}
\rho_{c}=\frac{\alpha (-1+m)A^{m}(1+z)^{mn}}{16\pi
(mA^{-1+m}(1+z)^{n(-1+m)}\alpha+\beta)}\left\{1+\frac{mn(2m-4)(1+z)^{4-2n}\left(3C_{1}-\frac{3k}{(1+z)^{2}}+\frac{A(1+z)^{-4+n}}{n-4}\right)}{A^{2}}\right\}
\end{equation}

\begin{equation}
\begin{array}{c}
  p_{c}=\frac{A(1+z)^{2n}}{48m\pi}\left\{-3A+6(-2+m)mn^{2}(1+n)^{2}(1+z)^{4-n}+\right.\\
  \left.+6mn(1+z)^{2-n}(-5+n+z+nz)\sqrt{C_{1}-\frac{k}{(1+z)^{2}}+\frac{A(1+z)^{-4+n}}{3(-4+n)}}\right\}
  \\
\end{array}
\end{equation}

Using (51), (52), (53) and (54) we get temperature $T$, squared
speed of sound $v_{s}^{2}$ and heat capacity $C_{v}$ as functions
of redshift $z$ in the following forms

\begin{equation}
T=\frac{\rho_{1}+p_{1}+\rho_{c}+p_{c}}{(1+z)^{3}S_{0}}
\end{equation}

\begin{equation}
v_{s}^{2}=\frac{\xi_{1}(z)}{\xi_{2}(z)}
\end{equation}
 where

\begin{equation}
\begin{array}{c}
  \xi_{1}(z)=\left[16(m(A(1+z)^{n})^{-1+m}\alpha+\beta)^{2}\times\right.\\
  \left.\left\{\frac{An(1+z)^{-1+4n}(-3A+6(-2+m)mn^{2}(1+n)^{2}(1+z)^{4-n}+6mn(1+z)^{2-n}(-5+n+z+nz)\sqrt{C_{1}-\frac{k}{(1+z)^{2}}+\frac{A(1+z)^{-4+n}}{3(-4+n)}})}{24m\pi}+\frac{A}{48m\pi}\times \right.\right.\\
  \left.\left.\left(-6(-2+m)m(-4+n)n^{2}(1+n)^{2}(1+z)^{3+n}+\frac{3mn(1+z)^{2+n}(-5+n+z+nz)\left(\frac{2k}{(1+z)^{3}}+\frac{1}{3}A(1+z)^{-5+n}\right)}{\sqrt{C_{1}-\frac{k}{(1+z)^{2}}+\frac{A(1+z)^{-4+n}}{3(-4+n)}}}+\right.\right.\right.\\
  \left.\left.\left.6mn(1+n)(1+z)^{2+n}\sqrt{C_{1}-\frac{k}{(1+z)^{2}}+\frac{A(1+z)^{-4+n}}{3(-4+n)}}-6m(-2+n)n(1+z)^{1+n}(-5+n+z+nz)\times\right.\right.\right.\\
  \left.\left.\left.\sqrt{C_{1}-\frac{k}{(1+z)^{2}}+\frac{A(1+z)^{-4+n}}{3(-4+n)}}\right)+\frac{3w_{m}(1+w_{m})(1+z)^{2+2n+3w_{m}}\rho_{m0}}{m(A(1+z)^{n})^{-1+m}\alpha+\beta}-\frac{A(-1+m)mnw_{m}(1+z)^{2+n+3w_{m}}(A(1+z)^{n})^{m}\alpha\rho_{m0}}{(m(A(1+z)^{n})^{m}\alpha+A(1+z)^{n}\beta)^{2}}\right\}\right]\\
\end{array}
\end{equation}

and

\begin{equation}
\begin{array}{c}
\xi_{2}(z)=\frac{1}{\pi(1+z)}\left[-(-1+m)mn(A(1+z)^{n})^{m}\left(1+\frac{2(-2+m)mn(1+z)^{4-2n}\left(3C_{1}-\frac{3k}{(1+z)^{2}}+\frac{A(1+z)^{-4+n}}{-4+n}\right)}{A^{2}}\right)\alpha\times\right.\\
\left.\{(-1+m)(A(1+z)^{n})^{-1+m}\alpha+m(A(1+z)^{n})^{-1+m}\alpha+\beta\}\right]+\\
\frac{2(-2+m)(-1+m)mn(1+z)^{3-2n}(A(1+z)^{n})^{m}\left(-6(-2+n)C_{1}+\frac{6k(4-5n+n^{2})(1+z)^{2}-An(1+z)^{n}}{(-4+n)(1+z)^{4}}\right)\alpha(m(A(1+z)^{n})^{-1+m}\alpha+\beta)}{A^{2}\pi}-\\
16(-1+m)mn(1+z)^{2+3w_{m}}(A(1+z)^{n})^{-1+m}\alpha\rho_{m0}+48(1+w_{m})(1+z)^{2+3w_{m}}(m(A(1+z)^{n})^{-1+m}\alpha+\beta)\rho_{m0}\\
\end{array}
\end{equation}

and
\begin{equation}
C_{v}=\frac{\zeta_{1}(z)}{\zeta_{2}(z)}
\end{equation}

where
\begin{equation}
\begin{array}{c}
\zeta_{1}(z)=\frac{(-1+m)mn(1+z)^{-1+n}(A(1+z)^{n})^{-1+m}\left\{A^{2}+2(-2+m)mn(1+z)^{4-2n}\left(3C_{1}-\frac{3k}{(1+z)^{2}}+\frac{A(1+z)^{-4+n}}{-4+n}\right)\right\}}{16A\pi(m(A(1+z)^{n})^{-1+m}\alpha+\beta)}\alpha\times\\
 \left\{1+\frac{\alpha(-1+m)(A(1+z)^{n})^{-1+m}}{(m(A(1+z)^{n})^{-1+m}\alpha+\beta)}\right\}+\\
+\frac{(-1+m)(A(1+z)^{n})^{m}\left\{2(-2+m)mn(1+z)^{4-2n}\left(\frac{6k}{(1+z)^{3}}+A(1+z)^{-5+n}\right)+2(-2+m)m(4-2n)n(1+z)^{3-2n}\left(3C_{1}-\frac{3k}{(1+z)^{2}}+\frac{A(1+z)^{-4+n}}{-4+n}\right)\right\}\alpha}{16\pi(m(A(1+z)^{n})^{-1+m}\alpha+\beta)A^{2}}\\
-\frac{A(-1+m)mn(1+z)^{-1+n+3(1+w_{m})}(A(1+z)^{n})^{-2+m}\alpha\rho_{m0}}{(m(A(1+z)^{n})^{-1+m}\alpha+\beta)^{2}}+\frac{3(1+w_{m})(1+z)^{-1+3(1+w_{m})}\rho_{m0}}{m(A(1+z)^{n})^{-1+m}\alpha+\beta}\\
\end{array}
\end{equation}

and
\begin{equation}
\begin{array}{c}
\zeta_{2}(z)=-\frac{3}{S_{0}(1+z)^{7}}\left[\frac{A(1+z)^{2n}\left(-3A+6(-2+m)mn^{2}(1+n)^{2}(1+z)^{4-n}+6mn(1+z)^{2-n}(-5+n+z+nz)\sqrt{C_{1}-\frac{k}{(1+z)^{2}}+\frac{A(1+z)^{-4+n}}{3(-4+n)}}\right)}{48m\pi}+\right.\\
\left.+\frac{(-1+m)(A(1+z)^{n})^{m}\left(A^{2}+2(-2+m)mn(1+z)^{4-2n}\left(3C_{1}-\frac{3k}{(1+z)^{2}}+\frac{A(1+z)^{-4+n}}{-4+n}\right)\right)\alpha}{16\pi A^{2}(m(A(1+z)^{n})^{-1+m}\alpha+\beta)}+\frac{(1+z)^{3(1+w_{m})}\rho_{m0}}{m(A(1+z)^{n})^{-1+m}\alpha+\beta}+\frac{w_{m}(1+z)^{3(1+w_{m})\rho_{m0}}}{m(A(1+z)^{n})^{-1+m}\alpha+\beta}\right]+\\
\frac{1}{S_{0}(1+z)^{6}}\times\left[\frac{A(1+z)^{2n}}{48m\pi}\left(-6(-2+m)m(-4+n)n^{2}(1+n)^{2}(1+z)^{3-n}+\frac{3mn(1+z)^{2-n}(-5+n+z+nz)\left(\frac{2k}{(1+z)^{3}}+\frac{1}{3}A(1+z)^{-5+n}\right)}{\sqrt{C_{1}-\frac{k}{(1+z)^{2}}+\frac{A(1+z)^{-4+n}}{3(-4+n)}}}+\right.\right.\\
\left.\left.6mn(1+z)^{1-n}\sqrt{C_{1}-\frac{k}{(1+z)^{2}}+\frac{A(1+z)^{-4+n}}{3(-4+n)}}((1+n)(1+z)-(-2+n)(-5+n+z+nz))\right)+\right.\\
\left.\frac{An(1+z)^{-1+2n}\left(-3A+6(-2+m)mn^{2}(1+n)^{2}(1+z)^{4-n}+6mn(1+z)^{2-n}(-5+n+z+nz)\sqrt{C_{1}-\frac{k}{(1+z)^{2}}+\frac{A(1+z)^{-4+n}}{3(-4+n)}}\right)}{24m\pi}-\right.\\
\left.\frac{(-1+m)mn(A(1+z)^{n})^{m}\alpha\left\{A^{2}+2(-2+m)mn(1+z)^{4-2n}\left(3C_{1}-\frac{3k}{(1+z)^{2}}+\frac{A(1+z)^{-4+n}}{-4+n}\right)\right\}}{16A^{2}(1+z)\pi(m(A(1+z)^{n})^{-1+m}\alpha+\beta)}\left(\frac{\alpha(-1+m)(A(1+z)^{n})^{-1+m}}{(m(A(1+z)^{n})^{-1+m}\alpha+\beta)}+1\right)\right.\\
\left.\frac{(-2+m)(-1+m)mn(1+z)^{3-2n}(A(1+z)^{n})^{m}\left(-6(-2+n)C_{1}+\frac{6k(4-5n+n^{2})(1+z)^{2}-An(1+z)^{n}}{(-4+n)(1+z)^{4}}\right)\alpha}{8A^{2}\pi (m(A(1+z)^{n})^{-1+m}\alpha+\beta)}+\frac{3(1+w_{m})(1+z)^{2+3w_{m}}\rho_{m0}}{m(A(1+z)^{n})^{-1+m}\alpha+\beta}+\right.\\
\left.\frac{3w_{m}(1+w_{m})(1+z)^{2+3w_{m}}\rho_{m0}}{m(A(1+z)^{n})^{-1+m}\alpha+\beta}-\frac{A(-1+m)mn(1+z)^{2+n+3w_{m}}(A(1+z)^{n})^{m}\alpha\rho_{m0}}{(m(A(1+z)^{n})^{m}\alpha+A(1+z)^{n}\beta)^{2}}(1+w_{m})\right]\\
\end{array}
\end{equation}

\begin{figure}
\includegraphics[height=2.8in]{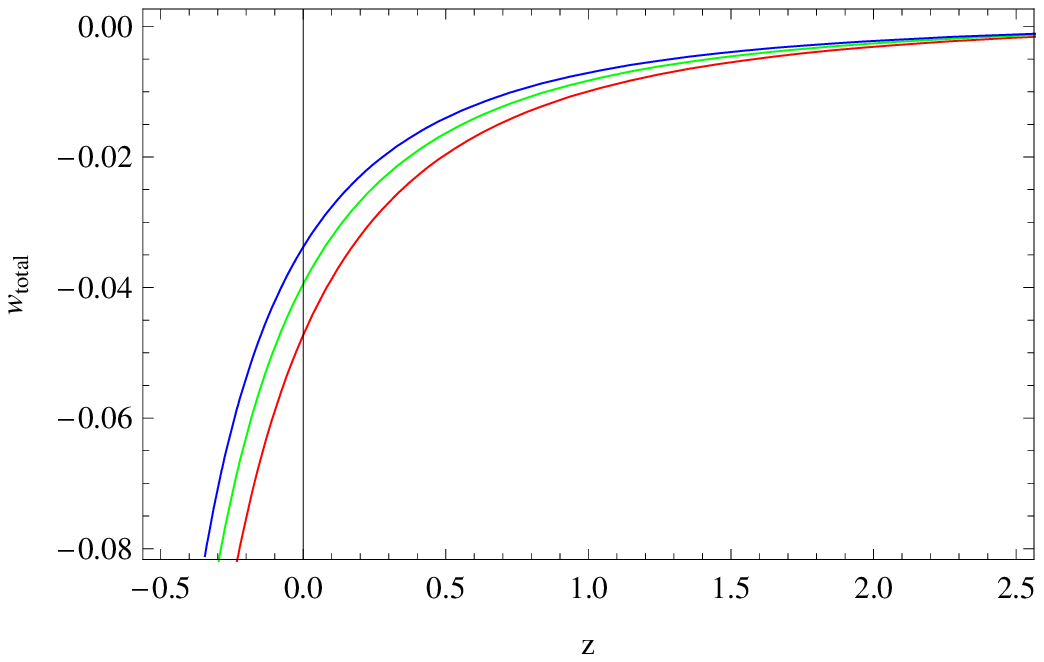}~~~\\
\vspace{1mm} ~~~~~~~~~~~~Fig.12\\
\vspace{6mm} Fig. 12 shows the evolution of the equation of state
parameter $w_{total}$ with the evolution of the universe in $f(R)$
gravity. We find that for $k=-1,~1,~0$ the equation of state
parameter $w_{total}>-1$. The indicates quintessence era.
Here,$\alpha=10.32,~\beta=10.01,~w_{m}=0.003,~m=12.3,~\rho_{m0}=0.23$
and the red, green and blue lines correspond to $k=-1,~1,~0$
respectively.

\vspace{6mm}

\end{figure}

The thermodynamic quantities expressed above are now plotted
against redshift $z$ to see their behavior with the evolution of
the universe. We proper choice of the parameters we plot all of
the quantities in figures 8, 9, and 10 respectively. In figure 8
we find the increasing behavior of the heat capacity with the
evolution of the universe. This behavior remains the same
irrespective of the curvature of the universe. From figure 9 we
see that as the universe in evolving, the temperature $T$ is
decreasing. Here also we get the same behavior for open, closed
and flat universes. It may be interpreted that the temperature of
the universe decreases as it expands under $f(R)$ gravity. In
figure 8 we plot the squared speed of sound $v_{s}^{2}$. Like the
temperature, $v_{s}^{2}$ is decreasing with the expansion of the
universe under $f(R)$ gravity. The choices of the parameters are
mentioned in the figure captions. Behavior of the equation of
state parameter $w_{total}$ is observed in figure 11. Throughout
the evolution of the universe $w_{total}>-1$. This indicates
quintessence like behavior of the equation of state parameter.
Therefore, we see that in $f(R)$ gravity, where we are considering
the coexistence of dark energy and dark matter with very small
pressure without interaction, the equation of state parameter
behaves like quintessence era. This holds true for flat, closed as
well as open universes. However, it also discerned that
$w_{total}$ is gradually increasing in the negative direction.\\

\section{\normalsize\bf{Discussions}}

In the present work, we have considered modified gravities as
Brans-Dicke, Ho$\check{\text r}$ava-Lifshitz and $f(R)$ gravities.
Various thermodynamic quantities like temperature, heat capacity
and squared speed of sound have been investigated for all the
gravity theories. In each case we have considered that the
universe is filled with dark matter and dark energy which are not
interacting. Prior to evaluating the thermodynamic quantities we
have studied the behaviors of the equation of state parameters. In
figure 1 we see that in the case of Brans-Dicke gravity theory,
the equation of state parameter is staying above $-1$ throughout
the evolution of the universe. This indicates that the equation of
state parameter is behaving like quintessence in this case
irrespective of the curvature of the universe.  In figure 5, where
we have plotted the equation of state parameter for the
Ho$\check{\text r}$ava-Lifshitz gravity, we see that the behavior
is like Brans-Dicke. It is staying above $-1$. However, at lower
redshifts the equation of state parameter is tending to $-1$.
However, it never crosses $-1$. Like Brans-Dicke, this behavior
remains the same for flat, open and closed universes. Similar
behavior of the equation of state parameter is discernible in
$f(R)$ gravity also. The evolution of the equation of state
parameter for $f(R)$ gravity has been presented in figure 12. From
the figures 3, 6 and 10 we see that the temperature  $T$ of the
universe is decreasing with evolution of the universe  in
Brans-Dicke, Ho$\check{\text r}$ava-Lifshitz and $f(R)$ gravities
respectively. This behavior remains the same in open, closed and
flat universes. From figures 2, 7 and 9 we find that the heat
capacity $C_{v}$ of the universe increases with the evolution of
the universe. Moreover, in all of the cases $C_{v}$ remains at the
positive level throughout the evolution of the universe. We have
also investigated the squared speed of sound $v_{s}^{2}$ in all of
the cases. From figure 4 we see that for closed universe
$v_{s}^{2}$ starts decreasing from redshift $-0.2$. However, up to
$z=-0.2$ it has gradually increased. In the cases of open and flat
universes, $v_{s}^{2}$ has an increasing behavior throughout the
evolution of the universe. From figure 8 we see that in
Ho$\check{\text r}$ava-Lifshitz gravity $v_{s}^{2}$  has a
decaying behavior throughout the evolution of the universe in the
case of open and closed universes. However, for flat $(k=0)$
universe, the squared speed of sound remains constant throughout
the evolution of the universe. From figure 11 we understand that
$v_{s}^{2}$ increases throughout the evolution of the universe
irrespective of the curvature of the universe.
\\

{\bf Acknowledgement:}\\

The authors wish to sincerely acknowledge the warm hospitality
provided by Inter-University Centre for Astronomy and Astrophysics
(IUCAA), Pune, India, where part of the work was carried out
during a scientific visit in January, 2011.
\\

{\bf References:}\\
\\
$[1]$ S. Carloni, M. Chaichian, S. Nojiri, S. D. Odintsov, M.
Oksanen and A. Tureanu, \emph{Phys. Rev. D} \textbf{82} (2010)
065020.\\
$[2]$ S. Nojiri and S. D. Odintsov, in \emph{Proceedings of the
42nd Karpacz Winter School of Theoretical Physics: Current
Mathematical Topics in Gravitation and Cosmology}, Ladek, Poland,
2006, \textbf{eConf C0602061}, 06 (2006).\\
$[3]$ S. Nojiri and S. D. Odintsov, \emph{Phys. Rev. D}
\textbf{68} (2003) 123512.\\
$[4]$ S. M. Carroll, A. De Felice, V. Duvvuri, D. A. Easson, M.
Trodden and M. S. Turner, \emph{Phys. Rev. D} \textbf{71} (2005)
063513.\\
$[5]$ S. Nojiri and S. D. Odintsov, \emph{Gen. Rel. Grav.}
\textbf{36} (2004) 1765.\\
$[6]$ A. D. Dolgova and M. Kawasaki, \emph{Phys. Lett. B}
\textbf{573} (2003) 1.\\
$[7]$ S. Capozziello, S. Nojiri, S. D. Odintsov, \emph{Phys. Lett.
B} \textbf{634} (2006) 93.\\
$[8]$ S. Nojiri and S. D. Odintsov, \emph{Phys. Lett. B} \textbf{576} (2003) 5.\\
$[9]$ S. Nojiri and S. D. Odintsov, \emph{Int. J. Geom. Meth. Mod. Phys.} \textbf{4} (2007) 115.\\
$[10]$ S. Nojiri and S. D. Odintsov, \emph{Journal of Physics:
Conference Series } \textbf{66} (2007) 012005.\\
$[11]$ P. Ho$\check{\text r}$ava, \emph{Phys. Rev. D} \textbf{79} (2009) 084008.\\
$[12]$ T. Faulkner, M. Tegmark, E. F. Bunn and Y. Mao,
\emph{Phys. Rev. D} \textbf{76} (2007) 063505.\\
$[13]$ E. O. Colgain and H. Yavartanoo, \emph{JHEP} \textbf{08}(2009) 021.\\
$[14]$ J. Kluson, \emph{JHEP} \textbf{07}(2010) 038.\\
$[15]$ M. Jamil, E. N. Saridakis and M. R. Setare, \emph{JCAP} \textbf{11} (2010) 032.\\
$[16]$ C. Brans and R. H. Dicke, \emph{Phys. Rev.} \textbf{124}
(1961) 925.\\
$[17]$ X. Chen and M. Kamionkowski,\emph{ Phys. Rev. D}
\textbf{60} (1999)104036.\\
$[18]$ A. Sheykhi, \emph{Phys. Lett. B} \textbf{681} (2009) 205.\\
$[19]$ A. Sheykhi, \emph{Phys. Rev. D} \textbf{81} (2010) 023525.\\
$[20]$ M. R. Setare, \emph{Phys. Lett. B} \textbf{644} (2007) 99.\\
$[21]$ N. Banerjee and D. Pavon, \emph{Phys. Lett. B} \textbf{647}
(2007)477.\\
$[22]$ L. Xu, W. Li and J. Lu,\emph{ Eur. Phys. J. C} \textbf{60} (2009) 135.\\
$[23]$ Y. Gong, \emph{Phys. Rev. D } \textbf{61} (2000) 043505.\\
$[24]$ B. Nayak and L. P. Singh, \emph{Mod. Phys. Lett. A} \textbf{24} (2009) 1785.\\
$[25]$ M. R. Setare and M. Jamil, \emph{Phys. Lett. B} \textbf{690} (2010) 1.\\
$[26]$ S. Das and N. Banerjee, \emph{Phys. Rev. D}\textbf{
78}(2008)043512.\\
$[27]$ T. Jacobson, \emph{Phys. Rev. Lett.} \textbf{75} (1995) 1260.\\
$[28]$ T. Padmanabhan, \emph{Class. Quant. Grav.} \textbf{19} (2002) 5387.\\
$[29]$ T. Padmanabhan, \emph{Physics Reports} \textbf{406} (2005) 49.\\
$[30]$ M. R. Setare and S. Shafei, \emph{JCAP} \textbf{09} (2006) 011\\
$[31]$ P. C. W. Davies, \emph{Class. Quant. Grav} \textbf{4}
(1987)L225.\\
$[32]$ H. M. Sadjadi, \emph{Phys. Rev. D} \textbf{73} (2006) 063525.\\
$[33]$ G. Izquierdo and D. Pavon, \emph{Phys. Lett. B} \textbf{639} (2006) 1.\\
$[34]$ H. Mohseni Sadjadi, \emph{Phys. Lett. B} \textbf{645} (2007) 108.\\
$[35]$ B. Wang, Y. Gong and E. Abdalla, \emph{Phys. Rev. D} \textbf{74} (2006) 083520.\\
$[36]$ W. Hu and I. Sawicki, \emph{Phys. Rev. D} \textbf{76} (2007) 064004.\\
$[37]$ S. Nojiri and S. D. Odintsov,\emph{ Phys. Lett. B} \textbf{576} (2003) 5.\\
$[38]$ L. Amendola, R. Gannouji, D. Polarski,
and S. Tsujikawa, \emph{Phys. Rev. D} \textbf{75} (2007) 083504.\\
$[39]$ H. M. Sadjadi, \emph{Phys. Rev. D} \textbf{76} (2007) 104024.\\
$[40]$ K. Bamba and C. -Q. Geng, \emph{Phys. Lett. B}\textbf{ 679 }(2009) 282.\\
$[41]$ S-F. Wu, B. Wang and G. -H. Yang, \emph{Nucl. Phys. B} \textbf{799} (2008) 330.\\
$[42]$ N. Mazumder and S. Chakraborty, \emph{Int. J. of Theor. Phys.} \textbf{50} (2011) 251.\\
$[43]$ S-F. Wu, B. Wang, G. -H. Yang and P. -M. Zhang, \emph{Class. Quant. Grav.} \textbf{25} (2008) 235018.\\
$[44]$ R-G. Cai and L. -M. Cao, \emph{Nucl. Phys. B} \textbf{785} (2007) 135.\\
$[45]$ A. Sheykhi, B. Wang and R. -G. Cai,  \emph{Nucl. Phys. B} \textbf{779} (2007) 1.\\
$[46]$ A. Sheykhi, \emph{JCAP} \textbf{05} (2009) 019.\\
$[47]$ A. Wang and Y. Wu, \emph{JCAP} \textbf{07}(2009) 012.\\
$[48]$ T. Nishioka, \emph{Class. Quantum Grav.} \textbf{26}(2009) 242001.\\
$[49]$ R-G. Cai and N. Ohta, \emph{Phys. Rev. D} \textbf{81}(2010) 084061 .\\
$[50]$ H. Kim and Y. Kim, \emph{Nuovo Cimento B}
\textbf{112}(1997) 329.\\
$[51]$ M. H. Dehghani, J. Pakravan and S. H. Hendi, \emph{Phys.
Rev. D} \textbf{74} (2006) 104014.\\
$[52]$ Y. Gong and A. Wang, \emph{Phys. Rev. Lett.} \textbf{99} (2007) 211301.\\
$[53]$ K. Bamba and C. -Q. Geng, \emph{Phys. Lett. B} \textbf{679} (2009) 282.\\
$[54]$ M. Akbar and R. -G. Cai, \emph{Phys. Lett. B} \textbf{648} (2007) 243.\\
$[55]$ H. M. Sadjadi, \emph{Phys. Rev. D} \textbf{76} (2007)104024.\\
$[56]$ S. Bhattacharya and U. Debnath, \emph{Int. J. Mod. Phys. D}
(2011) DOI: 10.1142/S0218271811019323.\\
$[57]$ E. J. Copeland, M. Sami and S. Tsujikawa, \emph{Int. J.
Mod. Phys. D}\textbf{ 15 }
(2006) 1753.\\
$[58]$ Y. Gong, B. Wang and A. Wang, \emph{Phys. Rev. D} \textbf{75} (2007) 123516.\\
$[59]$ E. Kiritsis and G. Kofinas, {\it Nucl. Phys. B} {\bf 821} (2009) 467.\\
$[60]$ M. Jamil, E. N. Saridakis and M.R. Setare, \emph{JCAP} \textbf{11} (2010) 032.\\
$[61]$ S. Nojiri and S. D. Odintsov, {\it Phys Lett B} {\bf
657} (2007) 238.\\
$[62]$ K. Bamba and C. -Q. Geng, {\it Phys. Lett. B} {\bf
679} (2009) 282.\\
\\

\end{document}